# Towards controllable Si-doping in oxide molecular beam epitaxy using a solid SiO source: Application to β-Ga$_2$O$_3$

A. Ardenghi[1*], O. Bierwagen[1*], A. Falkenstein[2], G. Hoffmann[1], J. Lähnemann[1], M. Martin[2], P. Mazzolini[3*]

[1] *Paul-Drude-Institut für Festkörperelektronik, Leibniz-Institut im Forschungsverbund Berlin e.V., Hausvogteiplatz 5-7, 10117 Berlin, Germany*

[2] *Institute of Physical Chemistry, RWTH Aachen University, D-52056 Aachen, Germany*

[3] *Department of Mathematical, Physical and Computer Sciences, University of Parma, Viale delle Scienze 7/A, 43124 Parma, Italy*

*Email: ardenghi@pdi-berlin.de - bierwagen@pdi-berlin.de - piero.mazzolini@unipr.it

The oxidation-related issues in controlling Si doping from the Si source material in oxide molecular beam epitaxy (MBE) is addressed by using its solid suboxide, SiO, as an alternative source material in a conventional effusion cell. Line-of-sight quadrupole mass spectrometry of the direct SiO-flux ($\Phi_{SiO}$) from the source at different temperatures ($T_{SiO}$) confirmed SiO molecules to sublime with an activation energy of 3.3eV. The $T_{SiO}$-dependent $\Phi_{SiO}$ was measured in vacuum before and after subjecting the source material to an O$_2$-background of $10^{-5}\ mbar$ (typical oxide MBE regime). The absence of a significant $\Phi_{SiO}$ difference indicates negligible source oxidation in molecular O$_2$. Mounted in an oxygen plasma-assisted MBE, Si-doped β-Ga$_2$O$_3$ layers were grown using this source. The $\Phi_{SiO}$ at the substrate was evaluated [from 2.9x10$^9$cm$^{-2}$s$^{-1}$ ($T_{SiO}$=700°C) to 5.5x10$^{13}$cm$^{-2}$s$^{-1}$ ($T_{SiO}$=1000°C)] and Si-concentration in the β-Ga$_2$O$_3$ layers measured by secondary ion mass spectrometry highlighting unprecedented control of continuous Si-doping for oxide MBE, *i.e.,* $N_{Si}$ from 4x10$^{17}$cm$^{-3}$ ($T_{SiO}$=700°C) up to 1.7x10$^{20}$cm$^{-3}$ ($T_{SiO}$=900°C). For a homoepitaxial β-Ga$_2$O$_3$ layer an Hall charge carrier concentration of 3x10$^{19}$cm$^{-3}$ in line with the provided $\Phi_{SiO}$ (T$_{SiO}$=800°C) is demonstrated. No SiO-incorporation difference was found between β-Ga$_2$O$_3$(010) layers homoepitaxially grown at 750°C and β-Ga$_2$O$_3$(-201) layers heteroepitaxially grown at 550°C. However, the presence of activated oxygen (plasma) resulted in partial source oxidation and related decrease of doping concentration (particularly at T$_{SiO}$<800°C) which has been tentatively explained with a simple model. Degassing the source at 1100°C reverted the oxidation. Concepts to reduce source oxidation during MBE-growth are referenced.

**THE MANUSCRIPT**

Due to a high variety of functional properties, metal oxides are rising in popularity as material systems for innovative optoelectronic devices.[1] Among metal oxides monoclinic β−Ga$_2$O$_3$ is one of the most interesting ones, and it has been particularly intensely investigated in the past decade.[2] In fact, due to a predicted breakdown field around 8 MV/cm as a consequence of its bandgap of about 4.8 eV, β−Ga$_2$O$_3$ is one of the most promising novel materials for power electronics.[3] Moreover, it can be grown from the melt[4–7] in turn allowing for β−Ga$_2$O$_3$ homoepitaxy of high quality layers.[3]

To obtain a broad range of devices for power electronic application, a thorough control on the electrical properties is needed.[8] Considering the lack of *p*-type doping for Ga$_2$O$_3$,[9] the main focus is on the group-IV elements in order to tune *n*-type doping. So far, Ge, Si and Sn have been already successfully employed as dopants in homoepitaxial layers grown with different technique such as Molecular Beam Epitaxy (MBE),[5,10–15] Metal Organic Chemical Vapor Deposition (MOCVD)[16–19] and Metal Organic Vapor Phase Epitaxy (MOVPE)[20,21].

Lany has theoretically shown[22] that Si can be considered as a better donor compared to Ge and Sn, since it is predicted to be the only truly shallow donor among them. From a survey of experimental data collected on semiconducting, homoepitaxial β−Ga$_2$O$_3$ layers,[8] the highest electron mobilities $\mu$ over a broad range of electron concentrations *n* are achieved with Si or Ge doping; the generally lower $\mu$ in Sn-doped films is likely related to a deep donor state identified for Sn.[12] Controlled Si doping in β-Ga$_2$O$_3$ has been already demonstrated in MOCVD but not in MBE. This is most likely linked to the two limitations of Si doping in MBE systems.

The first one is due to the unintentional incorporation of Si, probably related to the quartz cavity of the plasma source in oxygen plasma-assisted MBE (PAMBE)[23,24] and is limiting the low concentration side of the doping window ($N_{Si} < 10^{18}$ cm$^{-3}$).

The second one is due to instable Si-doping concentration in PAMBE-grown Ga$_2$O$_3$ layers, that depend on the oxygen background pressure rather than the Si source temperature as pointed out by Kalarickal *et al.*,[13] As underlying process, they identified the oxidation of the elemental Si source resulting in the source flux to consist of the SiO suboxide which has a higher vapor pressure than Si. Further oxidation of the Si surface into solid SiO$_2$ was identified by Krishnamoorthy *et al.*[15] to cause drastic flux reduction during layer growth due to the lower vapor pressure of SiO$_2$.

Indeed, the oxidation of the heated elemental source in an oxygen background into its volatile suboxide happens for many elements, including In, Ga, Ge, or Sn.[25] This can rule either the Ge- or Sn-doping of oxides (including in the Ga$_2$O$_3$ material system), or the growth rates for Ga$_2$O$_3$ during MBE deposition at relatively low Ga-source temperatures (*e.g.* <750 °C). Despite similar suboxides formation, the use of a Sn-metal source for Ga$_2$O$_3$ in PAMBE is more controllable with respect to a Si one,[12] but can involve segregation problems.[26] For Ge-doping of Ga$_2$O$_3$ from a Ge-source in PAMBE, an additional strong dependence of doping concentration on $T_g$ has been shown.[14] High concentrations (N$_{Ge}$ ≈ 10$^{20}$ cm$^{-3}$) required relatively low $T_g$ (600°C), potentially placing limits on the crystal quality of the deposited layers.

An efficient solution to control Si doping in oxide MBE while avoiding massive oxidation of the source, could be the use of an oxide or suboxide source material in the cell,[27] similar to the use of $SiO_2$, SnO, or mixtures of Sn and $SnO_2$ to produce a SnO flux,[26,28] or mixtures of Ga and $Ga_2O_3$ to produce a $Ga_2O$ flux.[28,29] Major advantages in the use of a suboxide (or mixed elemental+oxide) sources over an oxide one are the significantly lower cell temperatures required and the absence of parasitic oxygen formation.[28] On the other hand, controllable Si doping would benefit from a $SiO_2$ source that cannot oxidize further as SiO could. Based on the vapor pressure of $SiO_2$,[27] the geometry of our growth chamber, and the temperature limit of conventionally employed effusion cells (1200 °C), however, the Si concentration in layers deposited at a typical growth rate of 4.5 nm/min would be limited to $N_{Si} \leq 10^{19}$ cm$^{-3}$. For example, a high-temperature effusion cell run at $T_{SiO_2}=1400$ °C would be required to obtain $N_{Si} = 1 \times 10^{20}$ cm$^{-3}$ (considering that a GR of 1 Å/s in our MBE system correlate to a partial pressure of the source material in the effusion cell on the order of $10^{-3}$ mbar[28]).

In comparison, a SiO source could reach $N_{Si} = 1.7 \times 10^{20}$ cm$^{-3}$ at $T_{SiO}=900$ °C, which prompted us to characterize a solid SiO source for PAMBE growth with a focus on doping β-$Ga_2O_3$ layers. In particular, we demonstrate the possibility to obtain Si-doping concentrations on a wide range - from $4 \times 10^{17}$ cm$^{-3}$ (limited by the background Si concentration in the PAMBE deposited layers) up to $1.7 \times 10^{20}$ cm$^{-3}$ - in both homoepitaxial and heteroepitaxial β–$Ga_2O_3$ layers, with negligible dependence on the growth orientation and/or the growth temperature. For our study, an $Al_2O_3$ crucible loaded with 10 g of 3-6 mm SiO lumps (4N purity, Alfa Aesar) was placed in a conventional single-filament effusion cell.

At first, we characterized the direct flux from this SiO source using a quadrupole mass spectrometer (QMS) mounted line in sight to the cell inside a custom build system which is schematically shown in the inset of Figure 1 and described in Ref. [28]. The composition of the flux from the SiO source at a base pressure in the low $10^{-8}$ mbar regime is reported in Figure 1 (a) for a cell temperature of $T_{SiO} = 1200$ °C chosen to maximize the signal-to-noise ratio for the identification of the different desorbing species. Due to the natural abundance of stable isotopes of Si at 28 amu (92.23%), 29 amu (4.67%), and 30 amu (3.1%), the major signals shown in Figure 1 (a) can be clearly identified as SiO. A weaker Si signal is recorded and related to fragmentation inside the QMS.[28] To remove interference with residual $N_2$ (having the same mass per charge as $^{28}Si$) the measurements with closed cell shutter were subtracted from those with opened shutter. The same qualitative SiO spectra could be recorded down to $T_{SiO} = 880$ °C where the $^{44}SiO$ signal reaches the noise level.

To calculate the activation energy $E_a$ for SiO sublimation, the $^{44}SiO$ signal was recorded at $T_{SiO}$ ranging from 1200 °C to 800 °C in vacuum. The corresponding Arrhenius plot yields $E_a = 3.29$ eV (red line in (Figure 1 (b))) in good agreement with the $E_a = 3.41$ eV (black line in (Figure 1 (b))) extracted from the thermodynamic calculation of Adkison[27].

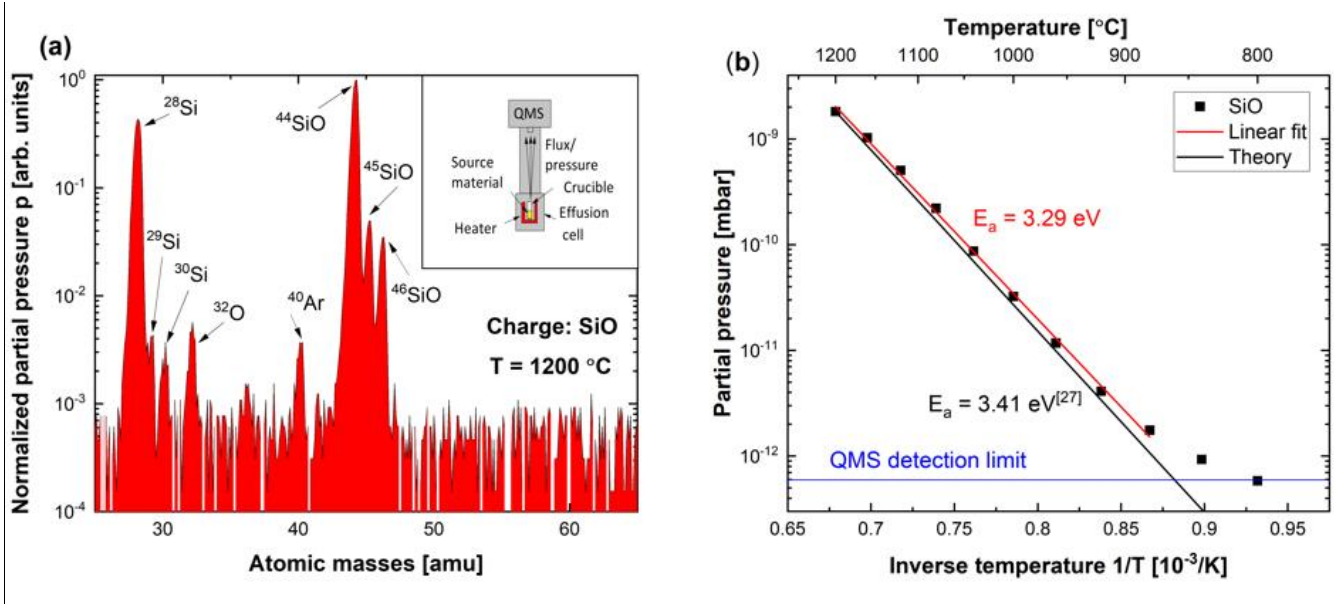

**Figure 1** (a) QMS spectrum (red) of the flux from the SiO cell at $T$=1200 °C in vacuum. Inset: Sketch of the system used for the measurement. (b) Arrhenius plot of the SiO signals collected in vacuum (black squares) for the experimental determination of the activation energy in comparison to the extrapolated activation energy from theory[27] (red and black lines, respectively).

To investigate potential source oxidation in molecular oxygen, the source was exposed to a controlled $O_2$ background of $p = 1\times10^{-5}$ mbar (typical for PAMBE growths) for periods of 30–180 minutes at various $T_{SiO}$. A potential surface oxidation to $SiO_2$ during this exposure is expected to reduce the SiO flux ($\Phi_{SiO}$) with respect to the untreated SiO source material. Figure 2 shows the comparison of the measured SiO partial pressure during the ramp up (performed in vacuum) from $T_{SiO}$=700 °C to 1200 °C before and after exposure of the source to an $O_2$ background pressure of $1\times10^{-5}$ mbar for 3 h at $T_{SiO}$=700 °C. The matching partial pressure recorded in both experiments indicate that the SiO source did not get oxidized. The initial (up to 15 min) higher values detected for the SiO partial pressure after the oxidation process (red dashed lines in Figure 2) are due to the higher background pressure caused by residual oxygen in the chamber after the treatment.

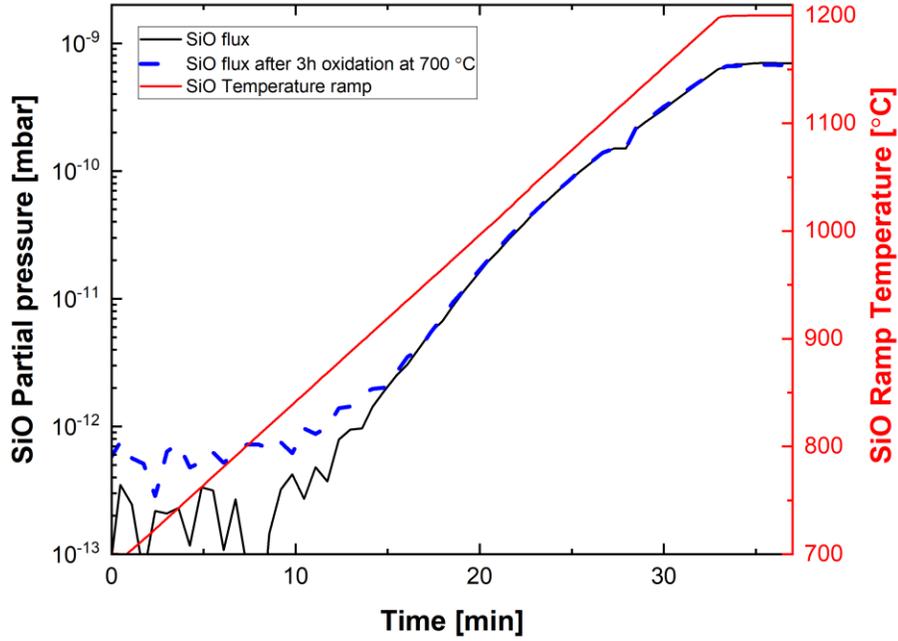

**Figure 2** Partial pressure from the SiO source for a ramp from 400 °C to 1200 °C before oxidation (black solid line) and after 3h with molecular oxygen background p = 10⁻⁵ mbar at $T_{SiO}$ = 700 °C (red dashed line). After the 3h at p = 10⁻⁵ mbar the O-flow was closed and the chamber was pumped down to p = 5 x 10⁻⁸ mbar. The source was then ramped up at 1200 °C.

Next, the crucible with the SiO source material was mounted in our PAMBE growth chamber for doping experiments. In order to predict the SiO-flux and consequential Si doping concentration in MBE deposited β-Ga$_2$O$_3$ layers as a function of the $T_{SiO}$, we extrapolated the corresponding $\Phi_{SiO}$ for $T_{SiO}$ =1000°C ($\Phi_{SiO}^{1000°C} = 5.5 \times 10^{13} cm^{-2} s^{-1}$) towards lower $T_{SiO}$. $\Phi_{SiO}^{1000°C}$ was extracted from the growth rate of an amorphous SiO$_2$ layer as described in the supplementary material.

$$\Phi_{SiO}(T) = \Phi_{SiO}^{1000°C} \, e^{[-E_a(1/kT - 1/k1000°C)]}. \qquad (1)$$

Eq. (1) provides this extrapolation, shown as the solid line in Figure 3, using the activation energy $E_a$ from the QMS experiment [red line in Figure 1 (b)]. Knowing the β-Ga$_2$O$_3$ growth rate - in our case a typical value is $GR_{Ga2O3}$ = 4.5 nm/min - it is possible to derive the expected Si doping concentration, e.g., $N_{Si}$ = 7.3x10²¹ cm⁻³ for $T_{SiO}$ = 1000 °C using:

$$N_{Si} = \frac{\Phi_{SiO}(T)}{GR_{Ga_2O_3}}. \qquad (2)$$

The predicted values obtained from Eq. (1) and Eq. (2) were used as our guideline to investigate different doping concentrations in β-Ga$_2$O$_3$ layers both on Al$_2$O$_3$(0001) and β-Ga$_2$O$_3$(010) substrates grown at $T_g$ of 550 °C and 750 °C, respectively (see supplementary material). Depth profiles of the Si dopant were obtained by time-of-flight secondary ion mass spectrometry (ToF-SIMS IV, iontof GmbH, Germany). For quantification, the relative sensitivity factor for silicon in a gallium oxide matrix was obtained by measuring an implantation standard of Si implanted into a nominally undoped β-Ga$_2$O$_3$ single crystal with a 10$^{14}$ cm$^{-2}$ fluence and an implantation energy of 80 keV.

In Figure 3 symbols represent the different $\phi_{SiO}$ at the substrate as function of the corresponding T$_{SiO}$, calculated as the product of $N_{Si}$ (measured by SIMS) and $GR_{Ga_2O_3}$ (obtained from SIMS profile) following Eq. 2. For reference, the corresponding volumetric Si concentrations are shown assuming a fixed growth rate of 4.5 nm/min (average GR of the deposited samples) on the right y-scale of Figure 3. The obtained broad range of Si doping concentrations from 3-4 x 10$^{17}$ cm$^{-3}$ (T$_{SiO}$ = 700 °C) up to 1.7 x 10$^{20}$ cm$^{-3}$ (T$_{SiO}$ = 900 °C) demonstrates the possibility to use a SiO source charge for continuous doping in β-Ga$_2$O$_3$ thin films controlled by the source temperature in PAMBE. Nonetheless, it was not possible to demonstrate the possibility of achieving doping concentrations below 3 x 10$^{17}$ cm$^{-3}$ due to an unintentional Si-doping in the nominally undoped β-Ga$_2$O$_3$ layers of around 2 x 10$^{17}$ cm$^{-3}$ detected by SIMS. We tentatively relate this unintentional Si incorporation to the plasma source as recently pointed out by Asel *et.al.*[24]

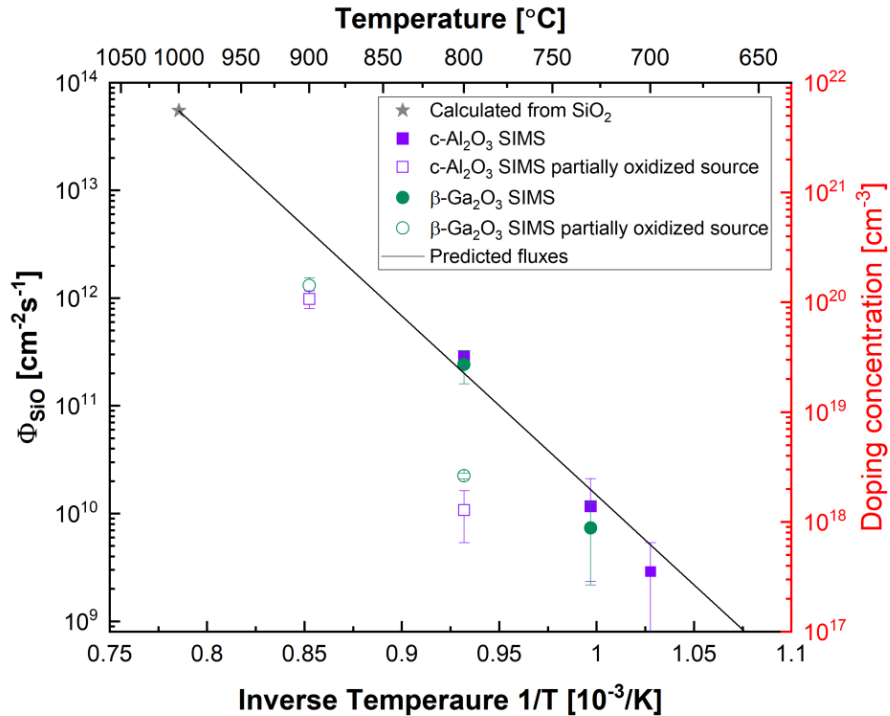

**Figure 3** SiO flux at the substrate as a function of the $T_{SiO}$; the right red scale shows the doping concentration considering a fixed β-Ga$_2$O$_3$ GR of 4.5 nm/min. Empty points are referred to layers grown with a partially oxidized source. For the filled ones, the SiO cell was ramp up at $T_{SiO}$ = 1100 °C before the deposition (dwell t = 30 min) in order to restore the SiO cell before the growth. The error bars represent the variation in the doping concentration along the layer thickness with respect to its average concentration. The solid black line refers to the extrapolated SiO flux calculated using Eq.1.

A close inspection of the Si doping profiles of all the deposited β-Ga$_2$O$_3$ films reveals a systematically decreasing Si concentration towards the surface, as exemplarily shown in Figure 4 (a). The respective highest and lowest Si concentrations from the begin and end of the doped layers are reflected by the corresponding error bars in Figure 3. Before the growth of the doped layers, the SiO cell was kept at the corresponding temperature for 15-60 min, in order to exclude the thermalization of the source as the possible origin of the slope. Consequently, we attribute the decreasing Si concentration in the doping profiles to progressive oxidation of the surface of the source material into SiO$_2$ during PAMBE growth. This finding is just apparently in contrast to the absence of source oxidation in molecular oxygen (see Figure 2), since it confirms the findings of Kalarickal et al.[13] who suggested active oxygen to play a major role for the oxidation of the Si source into SiO$_2$ in PAMBE. To further understand the impact of source oxidation, two different data set are provided in Figure 3. The empty symbols mark samples grown with a partially oxidized source, while the filled ones show data for samples where a high temperature cycle ($T_{SiO}$=1100 °C, 30-60 min) was carried out before the layer growth to degas and restore the SiO surface. At $T_{SiO}$=800 °C, the comparison between the two sets of data leads to a difference around one order of magnitude for the $\Phi_{SiO}$ (cf. Figure 3). Notwithstanding, the Si concentrations from the begin of the layers grown with a freshly degassed cell (upper error bar of filled symbols in Figure 3) follow the vapor pressure behavior of SiO, confirming the principal control of Si-doping by the source temperature rather than the background pressure. In addition, the matching incorporated SiO-flux concentration at the same $T_{SiO}$ both on (0001) Al$_2$O$_3$ ($T_g$ = 550°C) and β-Ga$_2$O$_3$ ($T_g$ = 750°C), strongly suggests *(i)* a negligible dependence of Si-incorporation on the $T_g$ as was instead observed for Ge,[14] and *(ii)* a negligible dependence on the β-Ga$_2$O$_3$ growth orientation [*i.e.*, (-201) and (010), respectively].

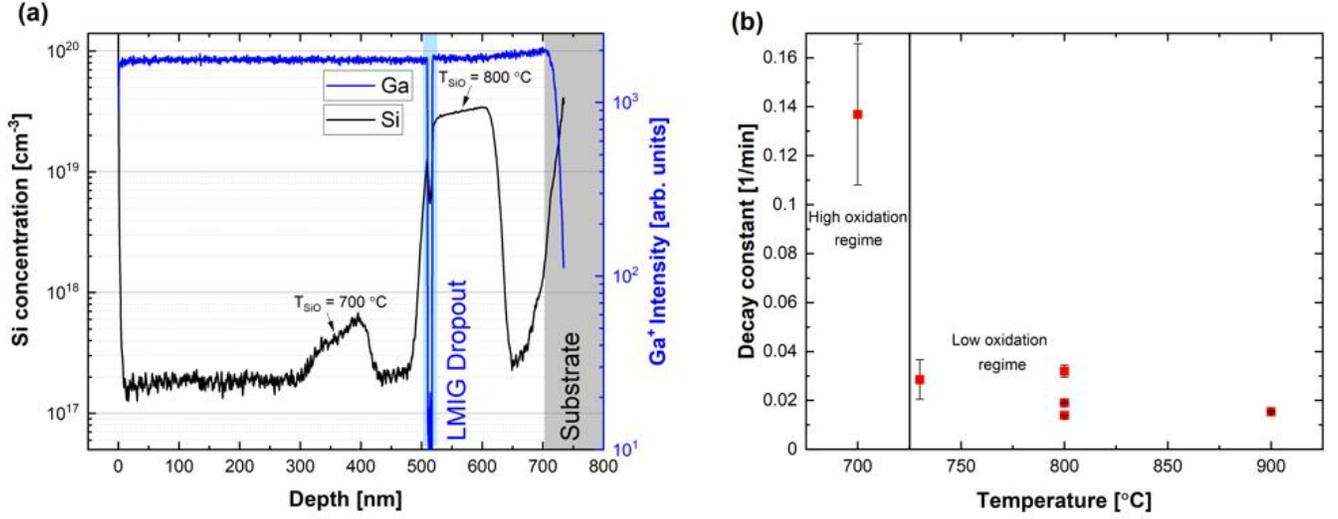

**Figure 4** (a) SIMS doping profile for a β-$Ga_2O_3$ sample grown with different $T_{SiO}$. The drop of the signal around 500 nm is caused by a dropout of the liquid metal ion gun (LMIG). A slope in the profile is visible, showing a steeper slope at lower $T_{SiO}$.
(b) Decay constant calculated from a linear fit of the SIMS profile in log scale. Before the fitting the Si background was subtracted and used to calculate the error bar. The error bars highlight how the background subtraction become important, *i.e.* how close the measured concentration is with respect to the background level (defined for nominally undoped layers).

Further inspection of Figure 4 (a) suggests the slope of the decreasing Si-concentration to depend on the SiO cell temperature, being steeper for a $T_{SiO}$ = 700 °C compared to the one at 800 °C. In a simple oxidation model (pictorially explained in the supplementary material, Fig. S2), the unoxidized area of the source material decreases with a rate constant λ. Consequently, an exponentially decreasing $Φ_{SiO}$ (which is proportional to this area) is expected during source oxidation:

$$Φ_{SiO}(t) = Φ_{SiO}(0)e^{-λ·t} \quad (3)$$

where $Φ_{SiO}(0)$ corresponds to the flux at *t*=0. Using Eq. 3, Eq. 2, and the relation of layer thickness *d* and corresponding growth time *t* ($d=GR_{Ga_2O_3} t$), we extracted λ from the slopes of the SIMS profiles d(ln$N_{Si}$)/d(*d*) as:

$$λ = GR_{Ga_2O_3} \, d(\ln N_{Si})/d(d) \quad (4)$$

Figure 4 (b) shows the extracted λ for layers doped at different $T_{SiO}$ and suggests two different oxidation regimes: *(i)* for a $T_{SiO}$ > 730 °C an almost flat doping profiles signify a low rate constant, *(ii)* for lower $T_{SiO}$, a high λ indicates non-stable doping throughout the layer thickness (*e.g.*, decrease in the doping concentration from 6.8x10$^{17}$ cm$^{-3}$ down to 2x10$^{17}$ cm$^{-3}$ in a deposition time *t*=15 min, *i.e.* over a 90 nm layer thickness). These data suggest lower $T_{SiO}$ (within the investigated range) to promote oxidation of the source, similar to the observations of Kim *et al.* for the Sr source oxidation.[30]

Finally, room temperature Van der Pauw-Hall measurements (details in supplementary material) were performed on a homoepitaxially grown sample with a 124 nm-thick Si-doped layer ($N_{Si} = 3 \times 10^{19}$ cm$^{-3}$). The extracted Hall electron concentration of $n = 3 \times 10^{19}$ cm$^{-3}$ with an electron mobility $\mu$ of 25 cm$^2$/Vs and a sheet resistance $R_s$ of 645 Ω/sq confirms effective Si-doping by our approach perfectly in line with the provided $\Phi_{SiO}$.

In conclusion, we have demonstrated the potential of a solid SiO source to provide a wide range of SiO fluxes controlled by the source temperature of a conventional effusion cell. Using this source, SiO fluxes at the substrate ranging from $5.5 \times 10^{13}$ ($T_{SiO}$=1000 °C) to $2.9 \times 10^{9}$ ($T_{SiO}$=700 °C) cm$^{-2}$s$^{-1}$ were used to grow a SiO$_2$ layer as well as continuously Si doping β-Ga$_2$O$_3$ layers with concentrations ranging from $1.7 \times 10^{20}$ ($T_{SiO}$=900 °C) cm$^{-3}$ to $3 \times 10^{17}$ ($T_{SiO}$=700 °C) cm$^{-3}$ inside an oxide PAMBE. Hall measurements of a homoepitaxial Si-doped β-Ga$_2$O$_3$ layer deposited with a $T_{SiO}$ corresponding to an expected $N_{Si} = 3 \times 10^{19}$ cm$^{-3}$ (800°C) revealed the same charge carrier density ($n = 3 \times 10^{19}$ cm$^{-3}$) indicating effective doping. The Si-incorporation did not depend on substrate temperature (550°C, 750°C) or the Ga$_2$O$_3$ growth orientation [(010), (-201)].

An oxidation of the SiO source, leading to a decreasing SiO flux, has been observed in active oxygen (plasma) but not in molecular oxygen at similar background pressures. The decreasing flux has been parametrized by a decay constant λ, extracted from the measured Si-doping profiles. Our initial data indicates a strong oxidation regime (λ>=0.14 min$^{-1}$) for relatively low $T_{SiO}$ (≤730 °C) and a milder one (λ< 0.015 min$^{-1}$) at higher $T_{SiO}$ (≥800 °C). The partial oxidation of the source can be reproducibly reverted by degassing at $T_{SiO}$ =1100 °C.

We believe that our approach can be also applied to O$_3$-MBE as well as for doping of other oxides, *e.g.* In$_2$O$_3$. Suggested ways to reduce the source oxidation during growth by oxygen PAMBE include *(i)* an increased $T_{SiO}$ or *(ii)* a decreased activated oxygen $p_{O2}$ at the SiO source. The *(i)* can be realized by a modified source geometry, *e.g.*, using an aperture on the crucible[31] or an increased source-to-substrate distance, [30] to necessitate a higher SiO vapor pressure (and thus higher $T_{SiO}$) inside the crucible to obtain the same SiO flux at the substrate. The *(ii)* can be realized using growth conditions that require less oxygen,[32] *e.g.*, metal-exchange catalyzed MBE (MEXCAT-MBE),[33–36] or a differentially pumped SiO source.[37]

**Acknowledgments:**

We would like to thank Duc Van Dinh for critically reading the manuscript, as well as Hans-Peter Schönherr, Claudia Herrmann, Carsten Stemmler, and Steffen Behnke for their technical support on the MBE and test chamber. This work was performed in the framework of GraFOx, a Leibniz-ScienceCampus, and was funded by Deutsche Forschungsgemeinschaft (DFG, German Research Foundation) - project number 446185170.

**Data Availability:**

The data that support the findings of this study are available from the corresponding author upon reasonable request

# Supplementary material to "Towards controllable Si-doping in oxide molecular beam epitaxy using a solid SiO source: Application to β-Ga2O3"

## Amorphous SiO₂ growth for SiO flux calibration

The amorphous SiO₂ layer was deposited on a 2" (0001) Al₂O₃ substrate at a $T_{SiO}$ = 1000 °C under conditions that provide full incorporation of SiO flux, i.e., a substrate temperature $T_{sub}$=100 °C, an oxygen flux of 1 standard cubic centimeter per minute (sccm), an RF plasma power $P_{RF}$ = 250W. The growth rate was calculated as ratio of film thickness (12 nm, measured ex-situ by X-ray reflectometry, Fig. S1) and growth time of 8 minutes. By using the mass density of amorphous SiO₂ ($\rho_{SiO_2}$ = 2.2 g/cm³) and the SiO₂ atomic mass ($M_{SiO_2}$ = 60 amu) we can calculate the density of SiO₂ molecules $N_{SiO_2}$ using the following formula:

$$N_{SiO_2} = \frac{\rho_{SiO_2} N_A}{M_{SiO_2}}, \quad (S1)$$

where $N_a$ is Avogadro's constant. Multiplying $N_{SiO_2}$ by $GR_{SiO_2}$, we obtained the silicon flux ($\Phi_{Si}$) at 1000°C for the SiO source:

$$\Phi_{SiO}^{1000°C} = GR_{SiO_2} N_{SiO_2} = 5.5 \cdot 10^{13} \, cm^{-2}. \quad (S2)$$

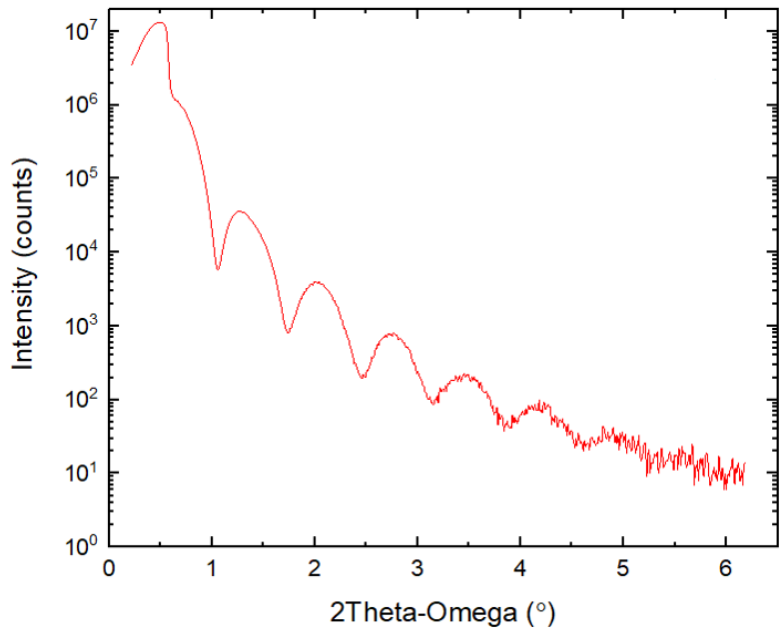

**Figure S1:** XRR curve for a 12nm thick amorphous SiO₂ layer.

## Growth details for β-Ga$_2$O$_3$

The growth of the homoepitaxial samples was performed on diced 5 x 5 mm$^2$ pieces of Fe-doped β-Ga$_2$O$_3$ semi-insulating substrates (Novel Crystal Technology) close to stoichiometric conditions (Ga beam equivalent pressure $BEP_{Ga}$ = 3.48 x 10$^{-7}$ mbar, $O$-$flux$ = 1 sccm, $P_{RF}$ = 250 W), at $T_g$ = 750 °C. The same growth conditions were applied for (-201)-oriented growth on 2" (0001) Al$_2$O$_3$ substrates with the only exception of a lower substrate temperature ($T_g$ = 550 °C) due to the orientation dependent GR in β-Ga$_2$O$_3$ by MBE,[33,38] and the initial growth of a nucleation layer (about 15-30 nm thick, $P_{RF}$ = 400 W).

## SiO source oxidation model

In order to physically understand the oxidation process than occurs at the SiO surface, when active oxygen is supplied, we analyzed the Si SIMS profiles as a function of the $T_{SiO}$ of several β-Ga$_2$O$_3$ layers. A simple oxidation model was then formulated to tentatively describe the oxidation process. Considering a fixed probability that a certain percentage of our source surface will further oxidize into SiO$_2$, as a function of $T_{SiO}$ and the oxygen background pressure in the MBE chamber, we expect an exponential decay of the accessible SiO surface, and hereby of the SiO flux. A simple sketch that explain the process is shown in Figure S2.

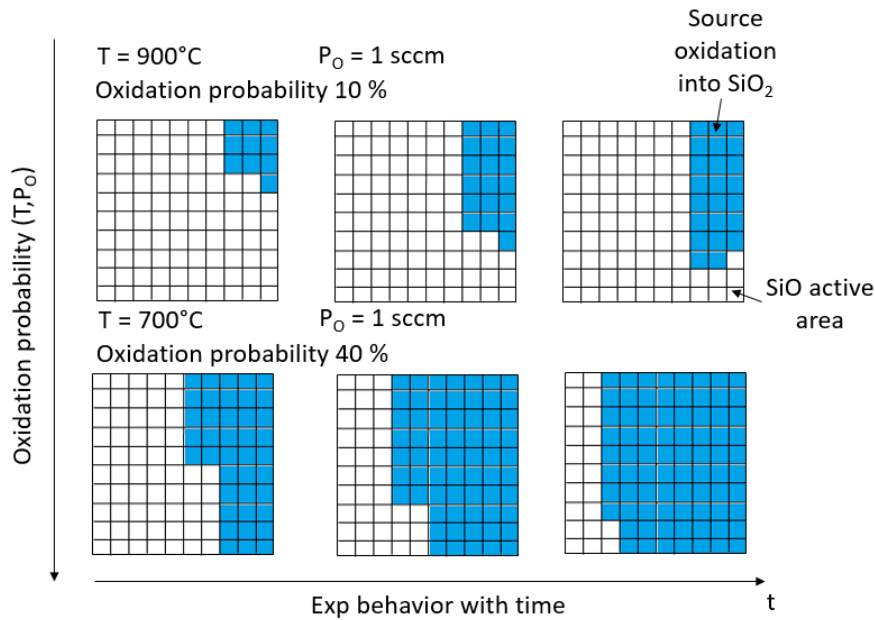

**Figure S2** Example of the oxidation process that may take place on our source as a function of $T_{SiO}$ and oxygen background pressure. In this case the oxygen pressure is fixed.

## Transport measurement

Van-der-Pauw-Hall measurements were performed at a current of 100 µA and a magnetic field of B = ±0.7 T oriented perpendicularly to the substrate surface. Prior to measurement, 20 nm Ti/ 20 nm Pt/ 100 nm Au circular contacts were deposited by electron-beam evaporation in the corners of the square sample through a shadow mask and confirmed to be ohmic without annealing by current-voltage measurements.